\documentclass[twocolumn,10pt]{revtex4-2}%
\usepackage[paperwidth=210mm,paperheight=297mm,centering,hmargin=2cm,vmargin=2.5cm]{geometry}
\usepackage{amsfonts}
\usepackage{amsmath}
\usepackage{amssymb}
\usepackage{bm}
\usepackage{graphicx}
\usepackage{color}
\usepackage{soul}
\usepackage{epsfig}%
\usepackage[dvipsnames]{xcolor} 
\definecolor{bleu_cite}{RGB}{0,0,255}
\usepackage[colorlinks=true,allcolors = black,citecolor=bleu_cite]{hyperref} 
\usepackage{siunitx}

\def\cl{\textcolor{black}}

\begin{document}

\title{Probing moiré excitons in MoSe$_2$/WSe$_2$ heterobilayers by combined micro-photoluminescence and lateral force microscopy}

\author{L. Caussou$^1$, H. Moutaabbid$^4$, M. Bernard$^1$, F. Margaillan$^1$, T. Taniguchi$^3$, K. Watanabe$^3$, C. Lagoin$^{1,2}$, F. Dubin$^{1,2}$ and V. Voliotis$^1$}
\affiliation{$^1$ Institut des Nanosciences de Paris, CNRS and Sorbonne Universit{\'e}, Paris, France}
\affiliation{$^2$ CRHEA, Université Côte d'Azur and CNRS, Valbonne, France }
\affiliation{$^3$ National Institute for Materials Science, Tsukuba, Ibaraki 305-0044, Japan}
\affiliation{$^4$ Institut de minéralogie, de physique des matériaux et de cosmochimie, CNRS and Sorbonne Universit\'e, Paris France}

\begin{abstract}
We study interlayer excitons in MoSe$_2$/WSe$_2$ heterobilayers, by combining lateral force microscopy and micro-photoluminescence spectroscopy. This allows us to correlate the spatial profile of the moiré superlattice with the distribution of optically active states accessible to interlayer excitons. In heterostructures where a few degrees twist angle is imposed between the MoSe$_2$ and WSe$_2$ crystallographic axes, we show that a continuous moiré lattice is realized across areas close to the optical diffraction limit. In such regions, the photoluminescence reduces to a few narrow-band lines only, energetically distributed consistently with the geometry of the moiré lattice. This correlation reveals that interlayer excitons explore a controlled periodic confinement, paving the way towards implementations of Bose-Hubbard models.
\end{abstract}

\maketitle

\textbf{\textit{Introduction }} Triggered by milestone studies with ultra-cold atoms \cite{Bloch_2008}, intense research efforts are dedicated to implement the Bose-Hubbard Hamiltonian in the solid state. This model describes strongly-interacting bosons confined in a lattice potential, yielding a rich variety of quantum solid phases such as Mott insulators. Semiconductor excitons, i.e. Coulomb-bound electrons and holes, are attractive candidates to build up such collective states \cite{lagoin_mott_2022}, provided that they thermalize to desired bath temperatures and explore a smooth lattice confinement, i.e. with ideally vanishing disorder. To meet these requirements, van der Waals assemblies of transition metal dichalcogenides (TMD) have recently opened a new avenue \cite{Jin_2023, Joon_2023,Gao_2024,Mak_2023}. 

To explore Bose-Hubbard physics, TMD heterobilayers are ideally suited. Indeed, these spontaneously  exhibit a triangular moiré superlattice, efficiently confining long-lived interlayer excitons, i.e. excitons made of electrons and holes each confined in a distinct monolayer \cite{Rivera_2018}. Interestingly, the moiré lattice is due to a periodic modulation of the stacking alignment between the atoms at the hetero-interface \cite{Mak_2022}. The period of the moiré potential is then tuned by varying the twist angle between the monolayer crystallographic axes. For example, it ranges from around 10 to 100 nm for MoSe$_2$ /WSe$_2$ bilayers \cite{berkelbach_theory_2013}. For a typical lattice depth between 20 and 100 meV \cite{wu_theory_2018,tran_evidence_2019,li_interlayer_2021}, repulsive dipolar interactions between interlayer excitons then theoretically stabilize Mott insulating phases, with for instance\cl{, }a single exciton in every lattice site, below a few tens of Kelvins \cite{fogler_high-temperature_2014,lagoin_key_2021,gotting_moire-bose-hubbard_2022}.

In angle-aligned MoSe$_2$/WSe$_2$, the moiré lattice has in principle the largest period, which is the most favorable situation to access Mott-like phases. In this quest, recent studies have shown that interlayer excitons radiate a photoluminescence (PL) made by an ensemble of narrow-band ($\sim$ 100 $\mu$eV) peaks, distributed over 10 to 30 meV \cite{seyler_signatures_2019,baek_highly_2020,brotons-gisbert_moire-trapped_2021,liu_signatures_2021}. This structure is similar to the emission profile of inhomogeneous quantum dot assemblies \cite{PhysRevLett.73.716}, pointing towards fluctuating energy minima of the moiré lattice sites. However, structured PL profiles are not systematic and rather sample dependent. Indeed, unrelated measurements revealed spectrally wide emissions in similar heterostacks \cite{tran_evidence_2019,hanbicki2018,ciarrocchi_polarization_2019,Holleitner_2023}. This discrepancy underlines the role of atomic reconstruction in angle-aligned bilayers \cite{weston_atomic_2020,enaldiev_stacking_2020,rosenberger_atomic_2020,Hogele_2023}. 
Precisely, weak intralayer strains provide an efficient channel to cancel the periodic modulation of atomic registries at the hetero-interface. Thereby the moiré lattice is suppressed and interlayer excitons effectively explore  two-dimensional domains where the luminescence is spectrally broader \cite{Ursulla_2024}.

In this Letter, we study MoSe$_2$/WSe$_2$ heterobilayers by combining atomic force microscopy (AFM) and micro-PL spectroscopy at 4 K. For samples in which interface adhesion is optimized by AFM ironing \cite{rosenberger_nano-squeegee_2018,palai_approaching_2023}, we show that a regular moiré lattice is realized only when the twist-angle between the monolayers crystallographic axis exceeds a few degrees. In that case, the moiré potential has a period reduced to around 10 nm, but most strikingly, we unveil that it is weakly varying across regions extending over the optical diffraction limit. In this situation, interlayer excitons are characterized by an "elementary" PL spectrum, consisting of only a few narrow-band lines ($\sim$1 meV large). Critically, we assign each emission line by only relying on the AFM measured lattice period. Our observations hence provide a key step towards controlled implementations of Bose-Hubbard models with moiré excitons.

\textbf{\textit{Nearly-aligned heterobilayers }} We first study angle-aligned MoSe$_2$/WSe$_2$ heterobilayers embedded in hexagonal boron nitride (hBN). The different layers are mechanically exfoliated from iodine CVT grown high quality bulk crystals \cite{hicham}, and stacked together using a PPC-based transfer technique \cite{lucille} before deposition on SiO$_2$ substrates. With a spatial resolution at the optical diffraction limit ($\sim$1 $\mu$m), \cl{electronic carriers are injected in the bilayer} through resonant laser-excitation of 1s intralayer excitons in WSe$_2$ \cl{(see Supplementary Information II)}. 
\cl{We then collect the PL emitted by interlayer excitons, which are unambiguously discriminated by their interaction with an external transverse electric field  (Supplementary Information II)}. In a first sample (denoted S1), Fig.1.a-b illustrates the typical strong spatial variations of the PL from interlayer excitons. On one hand, regions where the emission is the most intense result in unstructured and spectrally broad PL (light blue curve in Fig.1.b). Instead, multiple narrow-band peaks appear where the emission is ten times weaker (orange and yellow curves).

\begin{figure}
 \includegraphics[width=\columnwidth]{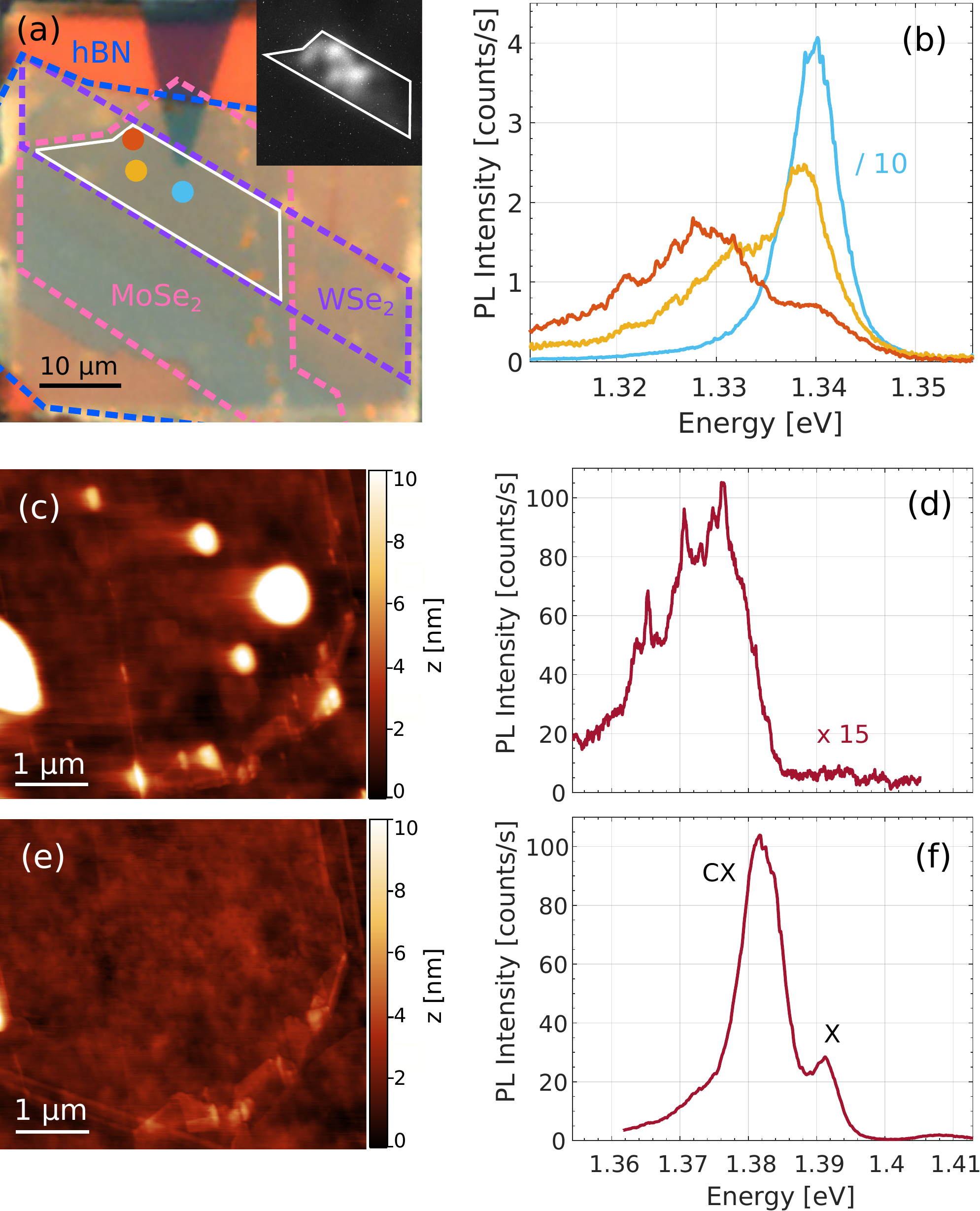}
 \caption{\textbf{Nearly-aligned MoSe$_2$/WSe$_2$ hetero-bilayers } (a) Optical microscope image of sample S1. Monolayers are each identified with separate colors while the inset displays the spatial map of the PL emission from interlayer excitons. (b) PL spectra for 3 different positions highlighted in (a). (c-e) Surface topography of sample S2 before (c) and after (e) AFM ironing through the 5 nm top hBN flake, with 1500 nN applied force. (d-f ) PL spectrum measured before (d) and after (f) AFM ironing at 4 K.}
\end{figure}

In Fig.1.b, the inhomogeneously distributed narrow band PL lines suggest that interlayer excitons are localized in the plane of the bilayer. However, accessing the process responsible for this spatial confinement is tedious, e.g. distinguishing moiré confinement \cl{from localisation due to defects or mechanical strain. Indeed, interface reconstruction, and related displacement of atomic registries at the hetero-interface \cite{weston_atomic_2020,Ursulla_2024,Hogele_2023}, provide efficient channels to possibly account for our observations.} Thus, a second sample (S2) was realized where an angle-aligned MoSe$_2$/WSe$_2$ bilayer was encapsulated with a thin h-BN top-layer (5 nm thickness). Hence, we correlate PL spectral profiles with the topography of the hetero-interface accessed through AFM. The surface topography is shown in Fig.1.c, where one notes that bubbles trapped in the hetero-stack yield a large device roughness, locally around 15 nm. In the same region, Fig.1.d evidences that the PL spectrum presents weak narrow-band lines. This correlation \cl{points towards strain-induced localisation of interlayer excitons, rather than their trapping in the moiré lattice sites.}

To further scrutinize the interplay between the interface topography and the PL spectral profile, sample S2 was ironed through the thin top h-BN, using an AFM tip in contact mode \cl{(Supplementary Information III)}. This approach has recently been shown to successfully improve the quality of embedded hetero-interfaces \cite{rosenberger_nano-squeegee_2018,palai_approaching_2023}. Figure 1.e directly confirms that the sample roughness is thus decreased to at most a few nm. In turn, Fig.1.f evidences that the PL intensity is increased by around fifteen-fold, as expected since the spatial overlap between the electron and hole wavefunction is then maximized. Moreover, the PL spectrum reduces to two main components, with around 4 meV spectral width and a separation about 8 meV. \cl{Additional measurements, realized in similar gated devices (Supplementary Information II), reveal that these emission lines are consistent with the radiative recombination of neutral (X) and charged (CX) interlayer excitons, delocalized in the plane of the bilayer in so-called $H_h^h$ stacking registries \cite{tran_evidence_2019}. This strongly suggests that excitons explore atomically reconstructed domains when interface adhesion is optimized by AFM-ironing. However, we expect that a canonical moiré confinement is possibly accessed at larger relative twist angles, i.e. in the regime where interface reconstruction becomes energetically unfavorable.}

\textbf{\textit{Small angle twisted heterobilayers }} \cl{Figure 2 quantifies the stacking alignment of a  third heterostructure, engineered with around 2° twist angle and AFM-ironed (sample S3). In Fig.2.a, we first verify that the device roughness is low (see also Supplementary Information III)}, particularly across the region investigated in the following (cyan rectangle). \cl{There, the stacking alignment is accessed through lateral force microscopy (LFM), which extracts the frictional properties between WSe$_2$ and MoSe$_2$ monolayers in AFM contact mode \cite{shi_moire_2017,huang_origin_2022}. It is worth noticing that in this case, no significant distorsions of the atomic registries at the hetero-interface are introduced (see Supplemental Information III)}. As LFM requires close contact between the AFM tip and the bilayer, the sample was fabricated using a very thin hBN top-layer (2 nm), allowing both LFM and PL spectroscopy. Figure 2.b presents the LFM image corresponding to the cyan rectangle in Fig.2.a. Across this area, a periodic modulation is directly observed. The fast Fourier transform of the LFM modulation (Fig.2.c) consists of hexagonally distributed peaks in reciprocal space. This pattern unambiguously reveals a moiré superlattice, with an average period around 9 nm, as expected for a twist angle around 2°. Figure 2.d further shows that the moiré potential is continuous and regular in real space.\vspace{.3cm}

\onecolumngrid

\centerline{\includegraphics[width=\textwidth]{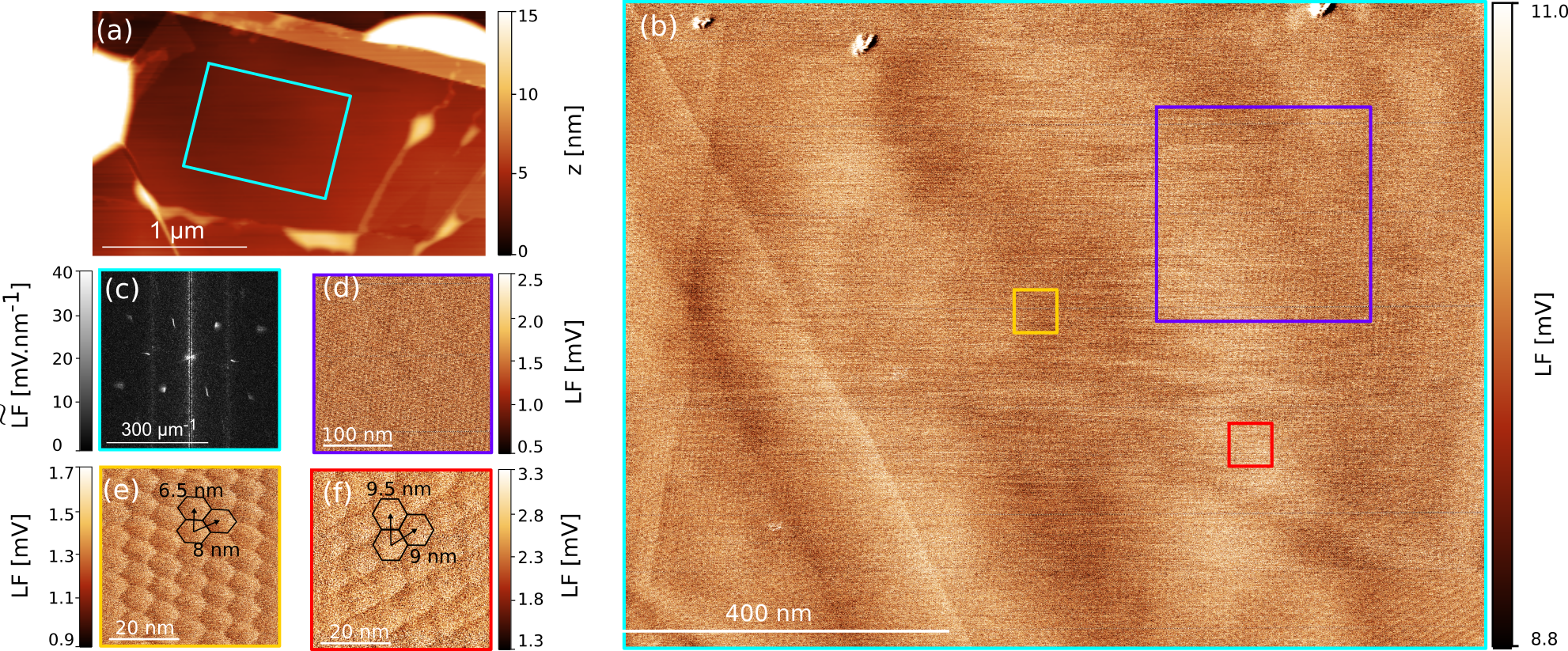}}
     \label{fig:enter-label}
FIG. 2. \textbf{Moiré superlattice for a 2° twist-angle} (a) AFM topography of a bilayer encapsulated with a 2 nm thin top hBN layer. (b) Lateral friction (LF) map measured in the region delimited by the cyan rectangle in (a), together with its fast Fourier transform $\tilde{\mathrm{LF}}$ (c). (d-f) High resolution LFM maps of the regions marked in panel (b) by the violet, yellow and red squares respectively.\\

\twocolumngrid

We performed high resolution LFM cartography in order to extract the spatial distortions of the moiré lattice geometry. Two limiting cases hence emerged: i) Fig.2.f shows that 100x100 nm$^2$ wide regions exhibit a regular moiré potential,  with a period $a_m$ that varies weakly between the symmetry axes, deviating by at most 0.5 nm from the 9 nm mean value obtained by averaging over the entire 1x1 $\mu$m$^2$ cyan region; ii) Fig.2.e shows that in other domains the moiré lattice is distorted, $a_m$ possibly varying fom 6.5 to 8 nm between two symmetry axes (yellow rectangle in Fig.2.b). The former value constitutes the lower bound of $a_m$ measured across the cyan rectangle in Fig.2.a. Additional LFM cartography (not shown) revealed that the largest value taken by $a_m$ is 11 nm. As discussed in the following, these fluctuations of the moiré lattice period are directly reflected by the PL spectrum.

Using the spatial period measured above for the moiré lattice, the energy distribution of the states accessible to interlayer excitons is directly extracted \cl{(see Supplementary Information IV)}. These levels are usually referred to as Wannier states (WS) and have energies only controlled by $a_m$ and by the lattice depth $V_0$ \cite{lagoin_key_2021,gotting_moire-bose-hubbard_2022}. The latter is theoretically expected around 30 meV \cite{wu_theory_2018,gotting_moire-bose-hubbard_2022}, and in the following we consider this value without any loss of generality. WS energies are represented in Fig.3.a, as a function of $a_m$. First, we note that a single state is accessible for $a_m\lesssim$9 nm, while for larger periods a second confined level emerges in the lattice sites, 15 to 20 meV energetically higher. \cl{Moreover, we note that WS energies decrease by over 5 meV when $a_m$ increase from 6 to 11 nm. The PL energy radiated by moiré excitons is then varied accordingly.}\\

\centerline{\includegraphics[width=\columnwidth]{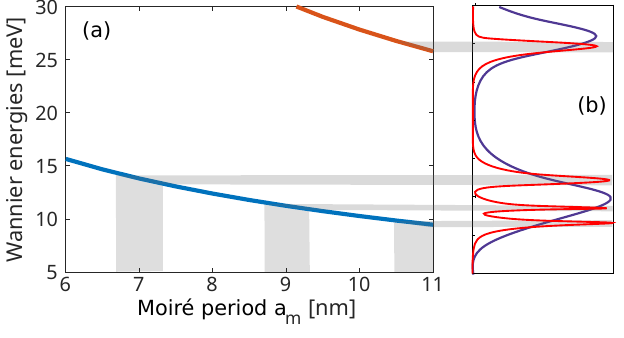}}
FIG. 3. \textbf{Moiré period and PL spectrum} (a) Wannier states energies as a function of the moiré period $a_m$, when the depth of the moiré potential is set to $V_0=$ 30 meV. Gray areas highlight the 3 spatial periods considered here, i.e. $a_m$= 7, 9 and 11 nm. \textbf{(b)} Schematic representation of the PL spectrum when $a_m$ equally takes all accessible values (purple curve), or when $a_m$ only equals to either 7, 9 or 11 nm (red curve). The spectral width ($\sim$ 1-2 meV in gray) is set by considering that $a_m$ varies by 1 nm around its accessible mean value.\\

\cl{Given the spatial distorsions of the moiré lattice observed by LFM, we expect the photoluminescence spectrum radiated by the cyan rectangle area to be representative of one of the two regimes: if $a_m$ equally takes all values accessible from 6 to 11 nm, then the spectrum should reduce to two broad lines separated by around 17 meV (purple curve in Fig.3.b). Each PL line is then centered at the WS energy set by the mean value of $a_m$ while the PL spectral width reflects the variance of $a_m$. In the contrary, if $a_m$ only takes a specific set of values then the PL spectrum should be spectrally structured.} The red curve in Fig.3.b illustrates an example obtained by considering that interlayer excitons are confined in domains where the moiré lattice has a period $a_m$ either equal to 7, 9 or 11 nm. The low energy component of the spectrum then splits into three narrow-band lines, each associated to the WS energy for the corresponding value of $a_m$. Interestingly, the energy separation between the PL lines quantifies the difference between the accessible lattice periods. Finally, a single emission is found at higher energies, due to the recombination of interlayer excitons occupying the second WS (in the lattice domain where $a_m$=11 nm).

\cl{Figure 4 quantifies the PL spectrum measured at the center of the AFM-ironed cyan area in Fig.2.a (see also Fig.SI.2 in the Supplementary Information). In the regime where the occupation of the moiré lattice is the lowest, i.e. for an excitation laser power density $P=0.02$ $\mu$W.$\mu m^{-2}$, Fig.4.a shows that the  profile reduces to a discrete number of peaks, with 6 low energy lines (labeled CX1 to CX3 and X1 to X3), and a higher energy one labeled X1’. Strikingly, the red curve in Fig.4.a shows that the energy positions of X1, X2 and X3 are quantitatively adjusted by considering PL emissions from interlayer excitons confined in domains where the moiré lattice period $a_m$ is either equal to 7, 9 or 11 nm. Given that these values align with LFM measurements, this agreement provides a strong indication for the radiative recombination of neutral interlayer excitons confined in the lowest WS of the moiré lattice sites. This conclusion is first supported by the emission energies of the additional peaks, CX1 to CX3. Indeed, these all lie around 7 meV below the corresponding X lines. This splitting matches closely the one measured between neutral and charged interlayer excitons (Fig.1.f), in good agreement with independent studies \cite{Xu_trion,brotons-gisbert_moire-trapped_2021}. This leads us deduce that a significant fraction of lattice sites is occupied by one exciton and an excess electron or hole. Furthermore, we note that the emission energy of X1' (Fig.4.a) is directly captured by considering the emission from 2$^{nd}$ WS of the domain where $a_m$=11 nm. Let us underline that this level is however weakly confined by the moiré lattice (Fig.3.a).}

Figure 4 highlights that the PL spectrum is quantitatively reproduced by only relying on the period of the moiré lattice measured by LFM (Fig.2). Nevertheless, to reach a good agreement we have adjusted the spectral widths of each emission line, between 1 and 2.5 meV. These magnitudes largely exceed our spectral resolution (100 $\mu$eV), but also the linewidth measured for excitons confined in artificial lattices of GaAs bilayers ($\mu$eV) \cite{lagoin_supersolid_2024}. Given the spatial distortions of the moiré lattice revealed by LFM, we attribute the spectral widths as a manifestation of the deviations of $a_m$ around its mean value. \cl{Indeed, from Fig.3.a, we note that WS energies shift by 1 to 2 meV when $a_m$ varies by around 0.5 to 1 nm (see gray areas). Such variations of the moiré period are expected from the measurements shown in Fig.2, thus effectively accounting for the spectral broadening of the PL lines in Fig.4.a (X and CX).}

\centerline{\includegraphics[width=\columnwidth]{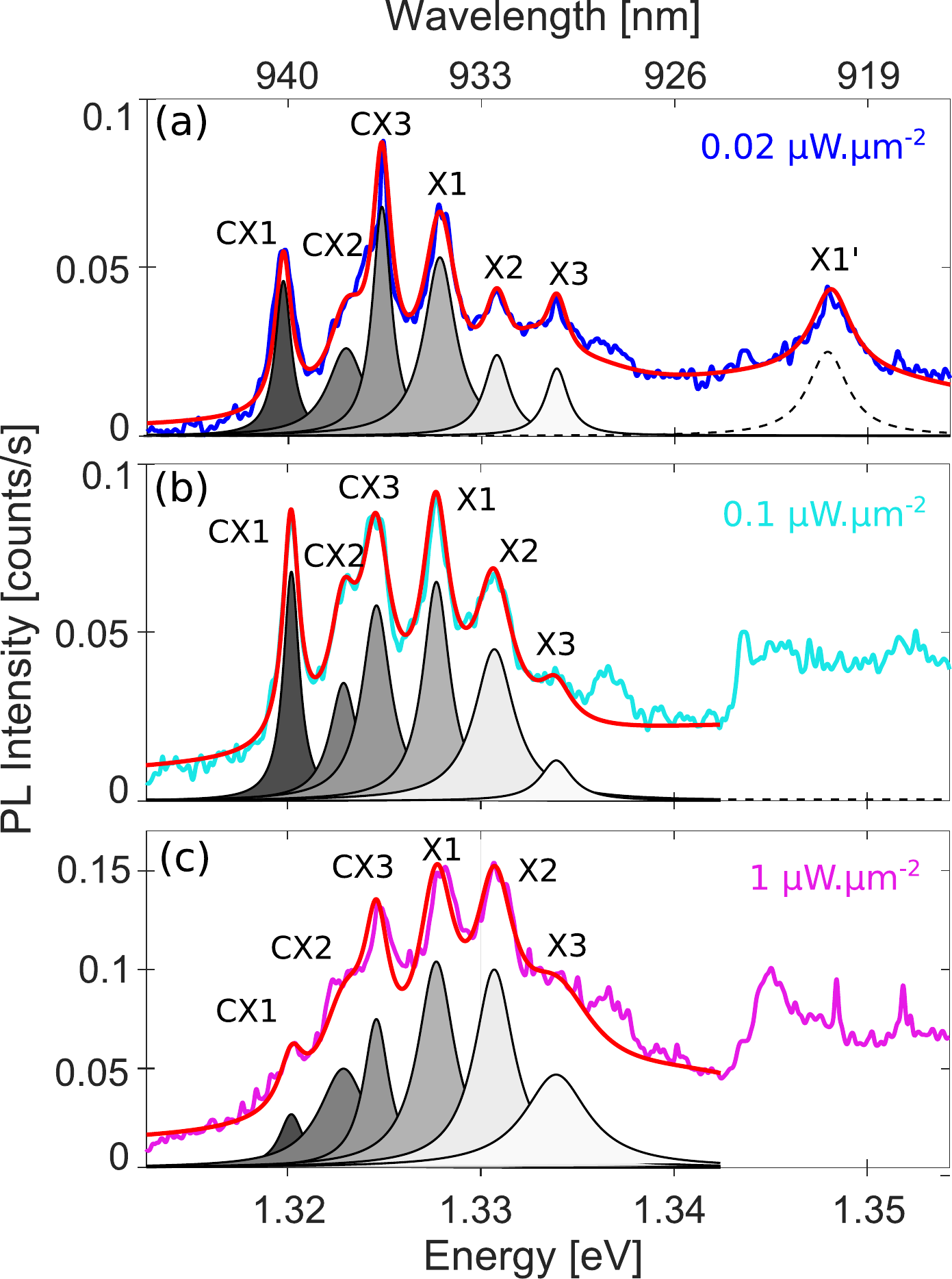}}
FIG. 4. \textbf{PL spectra for 2° twist-angle heterobilayer versus laser pump power.} PL spectrum at $P$= 0.02 $\mu W.\mu m^{-2}$ (a), $P$= 0.1 $\mu W.\mu m^{-2}$ (b), and  $P$= 1 $\mu W.\mu m^{-2}$ (c). Spectra were recorded in the region highlighted by the cyan rectangle in Fig.2.a, while model profiles (red curves) are computed by considering that the moiré lattice is made by domains where its period is either equal to 7, 9 and 11 nm (lines labeled X1-X3). Line X1' corresponds to the second accessible WS in the domain where $a_m=$ 11 nm. Lines labeled CX1 to CX3 are contributions from charged exciton complexes so that each CX line lies 7 meV below their neutral counterparts (X1 to X3, see text).

\cl{To further confirm exciton confinement in the moiré potential, we studied the variation of the PL spectrum while the exciton concentration is varied by the laser excitation power density $P$. To prevent laser-induced damage of our heterostructure, we restricted the latter to a maximum of 1 $\mu$W.$\mu$m$^{-2}$. In Figure 4.a to Fig.4.c, we compare the PL spectra measured for $P$ = 0.02, 0.1 and 1 $\mu$W.$\mu$m$^{-2}$. In this range, we quantitatively reproduce the spectral profiles by only varying the amplitude of each emission line (X1-X3 and CX1-CX3), but not their energy position. This behavior first reveals that interlayer excitons remain stably confined in the moiré lattice domains. Furthermore, it indicates that our experiments are bound to low lattice fillings where dipolar repulsions between excitons, confined in the same site \cite{Joon_2023} or in neighboring sites, play a negligible role. In this regime, interlayer excitons efficiently probe the energy distribution of accessible WS. Finally, we would like to underline that in Fig.4.b-c the spectra exhibit a step-like contribution at around 1.34 eV. We attribute this high-energy emission to the recombination of interlayer excitons that did not relax in the moiré potential, and instead remained in the continuum of states accessible above the lattice. Indeed, the energy  of the continuum starts around 15 meV above X1, as expected for $V_0\sim30$ meV (Fig.3.a). }

\textit{Conclusions } We have shown that lateral force microscopy and Photoluminescence spectroscopy provide complementary identifications to probe interlayer excitons confined in the moiré potential of MoSe$_2$/WSe$_2$ bilayers. LFM mapping allows one to extract the geometry of the moiré lattice across micrometric areas and thus deduce the corresponding spectral profile of the PL. In this way, we have evidenced that optimized nano-fabrication enables optical spectroscopy in regions where interlayer excitons are confined in a regular moiré lattice, with a period $a_m$ that only takes very few close values. LFM suggests that the small difference between the amplitudes taken by $a_m$ stems from local strains that distort of the moiré potential. Reducing further in-plane strains would then lead to model PL profiles, i.e. to the regime where the Bose-Hubbard phase diagram is mapped quantitatively. Moreover, we note that even in a moiré lattice with a sufficiently low level of disorder interlayer excitons can theoretically realize Mott insulating phases \cite{PhysRevA.81.063643,MOTT1968}.

\textit{Acknowledgements}
This work was funded by the French National Research Agency (ANR "IXTASE" Grant No. ANR-20-CE30-0032), while K.W. and T.T. acknowledge support from the JSPS (KAKENHI Grant Nos. 21H05233 and 23H02052) and the World Premier International Research Center Initiative, MEXT, Japan. Also, we thank M-L. Della Rocca, A. Reserbat-Plantey, S. Timpa, C. Morin and H. Cruguel for support during nano-fabrication and characterization. The authors declare no financial interest.

\onecolumngrid

\vspace{1cm}

\centerline{\large{\textbf{SUPPLEMENTARY INFORMATIONS}}}

\section{Samples fabrication}
We study MoSe2/WSe2 heterobilayers embedded in
hexagonal boron nitride (hBN), in either the nearly angle-aligned regime, or with a few degree twist angle. We use mechanical exfoliation from bulk crystals grown by iodine CVT \cite{hicham}, and a PPC-based transfer technique, to fabricate heterostructures. The twist angle between symmetry axes of the stacked monolayers is imposed by selecting flakes with at least one sharp edge along a well-defined crystallographic ax orientation, or two sides forming an angle of a multiple of 60°. Monolayers are thus stacked with an angular precision around 1$^{\circ}$.

\section{Micro-photoluminescence at 4 K}

\centerline{\includegraphics[width=0.75\columnwidth]{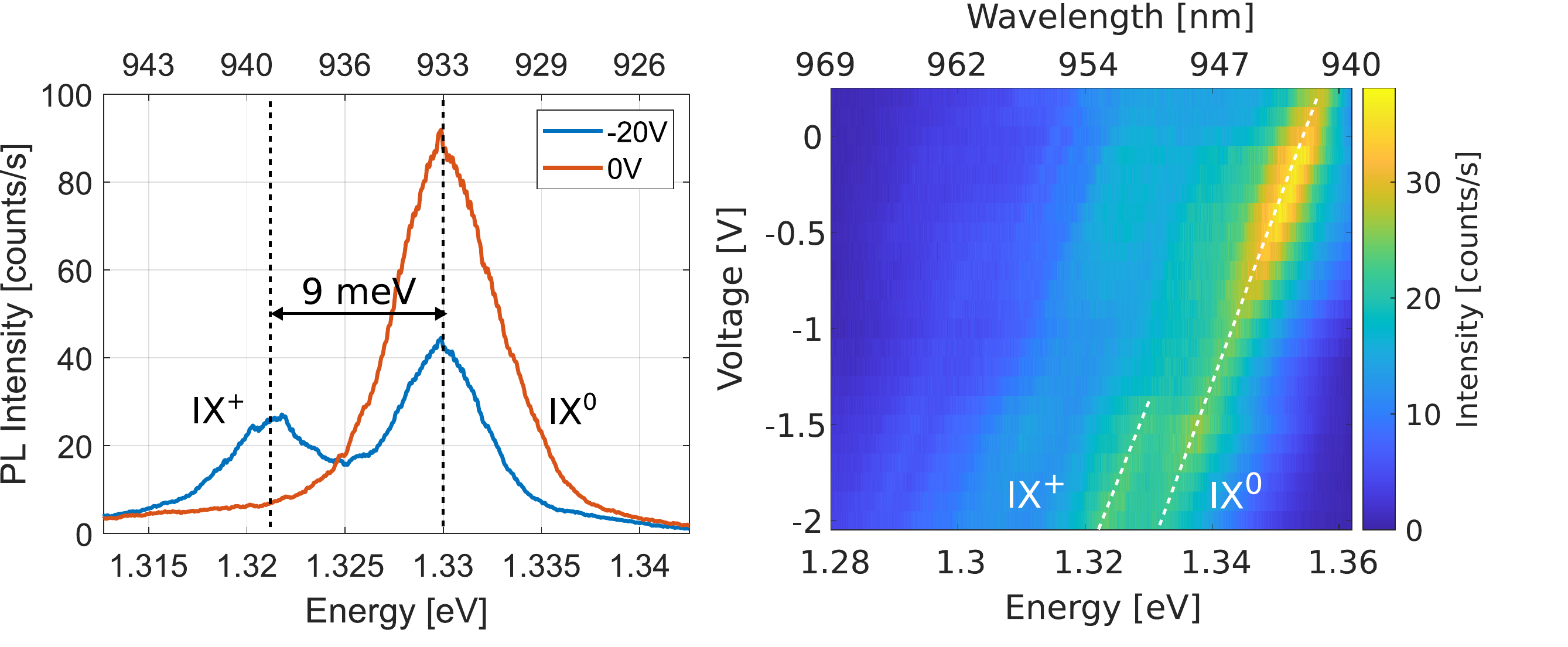}}
{\textbf{Fig. SI.1. Gate dependent PL measurement} (a) PL spectra when the top gate voltage is set at 0 V and -2 V, while the bottom gate is grounded. At 0 V, a single line marks the recombination of neutral interlayer excitons (IX$^0$), while at lower voltage a second line appear at 9 meV lower energy, corresponding to charge excitons (IX$^+$). (b) PL energy as a function of the top gate voltage. Both neutral and charged interlayer exciton lines show a linear shift marking the excitons permanent electric dipole.}\vspace{.5cm}

All the experiments presented here are performed at 4 K, in a Helium flow optical cryostat. Electron-hole pairs are photo-created in the WSe$_2$ layer, by pumping with a laser at 732 nm, at resonance with the intralayer band-to-band transition. Interlayer excitons are formed once electrons and holes have tunneled towards their
minimum energy states, lying in distinct layers, i.e. WSe$_2$ for holes and MoSe$_2$ for electrons. Thus, bilayer excitons have a fixed permanent electric dipole, given by the interlayer spatial separation. Moreover, the laser excitation is focused on the sample by a microscope objective with x50 magnification, and a numerical aperture of 0.5, which yields a gaussian spot with a full width at half maximum of 1 $\mu$m. The photoluminescence (PL) spectrum is dispersed with a 1200 lines/mm, leading to a spectral resolution of 0.3 nm.

Samples studied in the core of the manuscript were not electrically gated, since combining LFM and micro-PL requires a very thin ($\sim$2 nm) top hBN layer. Nevertheless, we also studied gated samples, this time with a thick h-BN top layer (around 20 nm) to prevent electrical breakdown. By relying on top and bottom graphene gates, we thus applied a uniform electric field, perpendicular to the bilayer. Note that we precisely kept the bottom gate grounded, while the potential applied on the top gate was varied.

Figure SI.1 shows a PL intensity map as a function of the applied top gate voltage. At zero bias, we observe a single emission line, corresponding to neutral indirect exciton (IX$^0$). When decreasing the gate voltage a second emission emerges at lower energy, around 9 meV below the IX$^0$ energy. It manifests the energy of charged excitons (IX$^+$), since for non-vanishing applied bias we electrically dope the bilayer, the bottom gate being grounded. Furthermore, both resonances experience a linear shift $\Delta E$ (dotted lines), as expected for interlayer excitons with an electric dipole moment $e\cdot d$ so that $\Delta E=edF$, $F$ denoting the applied electric field and $e$ the electron charge. From the optical contrast of the device, we estimate that top and bottom h-BN layers have thicknesses equal to (15 ± 5) nm and (20 ± 5) nm respectively. We then deduce $d = e(0.78 \pm 0.25)$ nm. This value matches well the one expected from the interlayer distance (0.65 nm). These observations are thus consistent with the interlayer nature of the excitonic emission observed around 1.33 eV in our studies.

In Fig.4 of the main text, we focus on the center of the ironed area (Fig.2), where we measured the moiré lattice with LFM. Figure SI.2 shows PL measurements at 4 K at different positions inside and around this region. We first note that all spectra exhibit several narrow-band peaks in the IX energy range, except the last one (number 5) taken on the yellow position outside of the moiré area. We also observe that the number of peaks and their energy vary spatially, pointing out the inhomogeneities of the moiré superlattice at the micrometric scale. As for the spectrum at position 5, it only presents broad lines with very weak intensity, consistent with a vanishing moiré lattice depth or with a shallow moiré superlattice, as a result of a weak interlayer coupling. Accordingly, the PL is spectrally broad. On the other hand, at the center of the ironed area, we note that spectra at position 1 and 2 are similar, as expected for weak spatial deformations of the moiré lattice. In the main text, we focus on the PL spectra taken at position 1.

\centerline{ \includegraphics[width=\columnwidth]{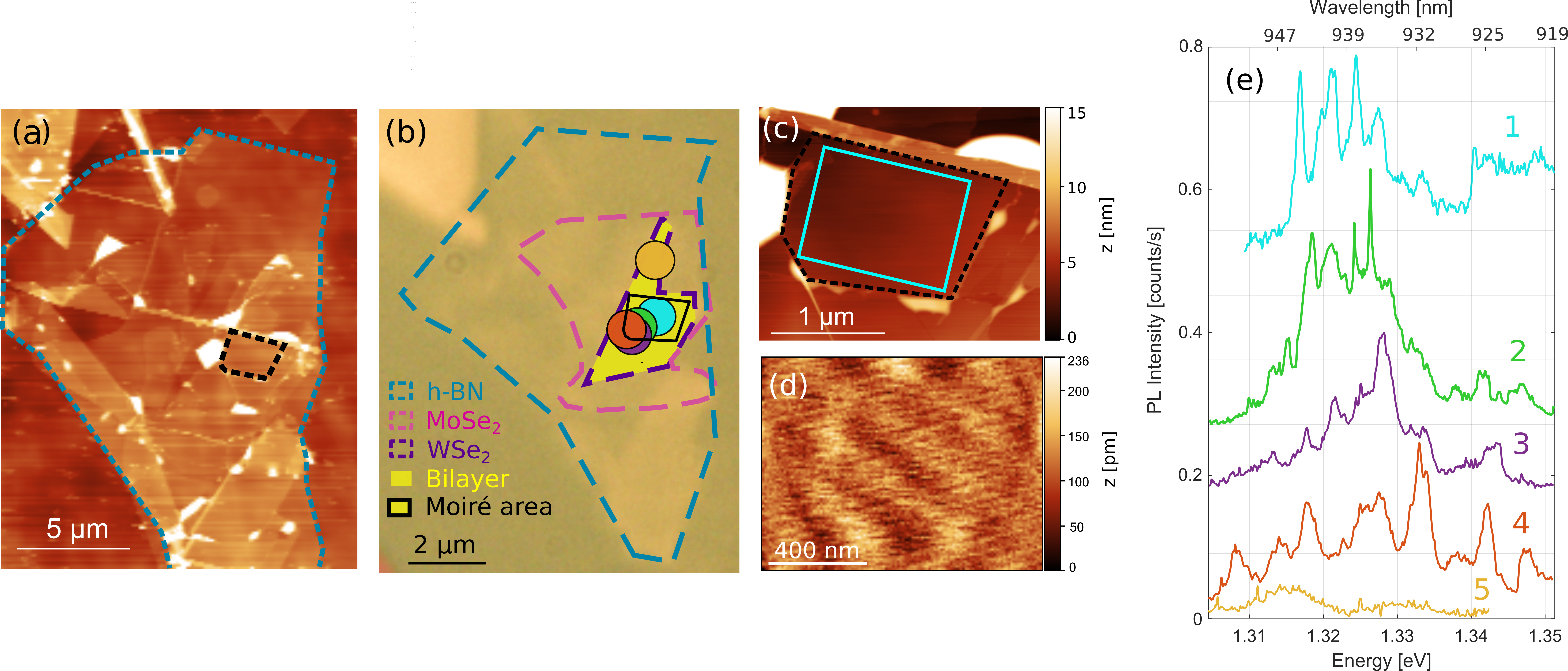}}
{\textbf{Fig. SI.2. Position dependent PL spectra.} (a) AFM image corresponding to ironed sample 3.
 (b) Optical microscope image showing the detailed heterostructure.
(c) Zoom in AFM image of the ironed area studied in Fig.2 and Fig.4 (cyan rectangle).
(d) AFM topography map measured at the center of the region highlighted by the cyan rectangle (same as Fig.2.a) showing that the surface roughness is reduced to less than 200 pm.
(e) PL spectra taken at the different positions indicated by colored dots in (b). In all figures, the black dotted contour corresponds to the studied moiré area.}

\section{Atomic Force Microscopy and Lateral Force Microscopy}
\textbf{AFM imaging:} the surface topography of our heterostructures, as shown in Fig.1.c and Fig.1.e, is measured with an atomic force microscope (AFM) NX20 from Park Systems in tapping mode with an instrumental resolution of 0.02 nm.
\\

\textbf{AFM ironing:} to mechanically push bubbles away from areas of interest, we use the AFM tip in contact mode. FMV-Pt probes from Bruker were used, as their tip diameter is larger, allowing for more efficient ironing. After a first scan with a low applied force of 7.5 nN, the force is gradually increased until it is sufficient to observe a displacement of the undesired blisters. This ensures that we apply the minimal force required in order to avoid damaging the sample. For large bubbles, repeating the scans may be necessary, but usually a single scan is sufficient to remove contaminants.
\\

\textbf{LFM imaging of moiré superlattices:} we use Lateral Force Microscopy (LFM) to record the in-plane lateral deflection of the tip (instead of the vertical deflection for topography), see Fig.SI.3. This technique is typically used to detect frictional properties of a material surface. For these measurements, FMV-Pt probes were used, as they have low spring constant and therefore give rise to a stronger lateral torsion. As LFM records the surface friction, it can not evidence the moiré superlattice if the TMD bilayer is covered with a thick h-BN flake. Therefore, in this study we use thin h-BN top layers (2 nm) in order to allow for joint PL and LFM measurements. Moreover, a force of 1000 nN is used with scan rates close to 1 Hz. Fig.SI.3 shows that while the topography image (a) does not show any variation, except for a small linear defect, the lateral force image (b) clearly resolves an hexagonal moiré superlattice, with 8 nm period, corresponding to a twist angle around 2.3°.\vspace{0.3cm}

 \centerline{\includegraphics[width=0.5\columnwidth]{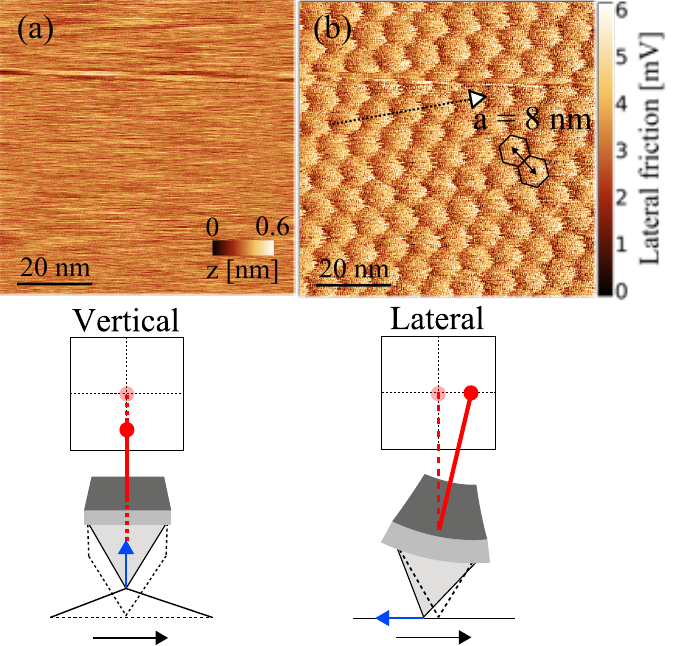}}
 {\textbf{Fig. SI.3. AFM and LFM scanning modes.} AFM (a) and LFM (b) images of the same area on the sample with their working principles below. While the AFM measures the vertical deflection and shows no topographic structure, LFM measures the lateral friction and allows to observe the hexagonal moiré pattern with 8 nm period.}\vspace{0.3cm}

Fig SI.4 compares LFM scans realized for forward (a) and backward (b) directions, after AFM ironing. The lateral force being opposite, the color scale is reversed, but no other difference is noticed. In particular, Fig SI.4 shows that the AFM ironing and/or an LFM scan do not modify the geometry of the moiré lattice.\vspace{0.3cm}

\centerline{\includegraphics[width=0.5\columnwidth]{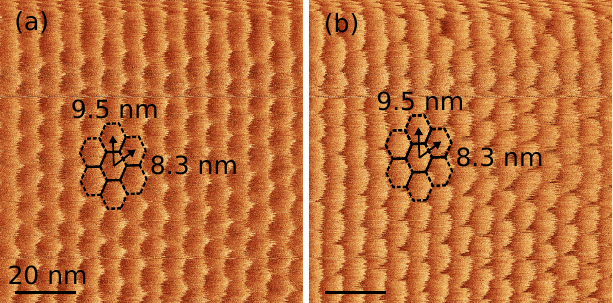}}
\textbf{Fig. SI.4. Role of LFM scanning direction.} LFM forward (a) and backward (b) image showing no distortion of the moiré pattern due to LFM tip applied force. \\

\centerline{ \includegraphics[width=0.5\columnwidth]{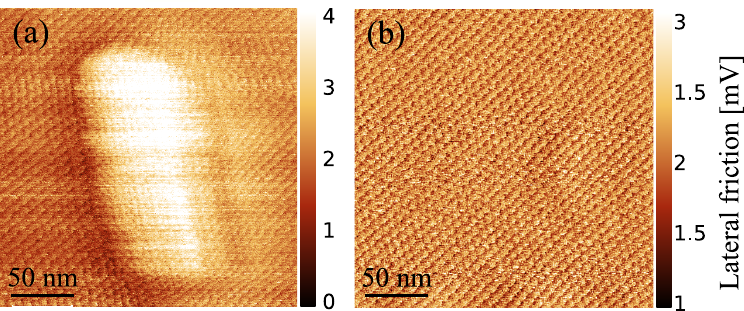}}
\textbf{Fig.SI.5. AFM ironing.} LFM images before (a) and after (b) ironing of the sample. While in (a) a 5 nm thick blister is visible, it is properly removed by ironing unveiling a regular moiré pattern. \\

Figure SI.5 shows LFM images on the same area of sample S2 before and after ironing a 5 nm thick blister. In the first image, the moiré is deformed by the blister, but still visible, which highlights the strength of the interlayer adhesion. Ironing is performed by increasing the setpoint from 0.01 V (7.5 nN) to 0.05 V. As a result, the deformation is pushed away, leaving a clean moiré superlattice. Importantly, this measurement reveals that ironing with low applied forces does not destroy the moiré lattice structure.

\section{Bose Hubbard model for interlayer excitons}
\subsection{Lattice potential}
Let us approximate the moiré potential by considering a one dimensional lattice $V$, with an amplitude $V_0$ and a sinusoidal profile
\begin{equation}
V(x)=V_0 \sin^2{(Qx)}
\end{equation}
with $Q=\pi/a_m$, where $a_m$ is the moiré lattice period.
Note that $a_m$ is controlled by the twist angle, since $a_m\approx a_0/\sqrt{\theta^2+\delta^2}$, with $\delta$ the lattice mismatch between WSe$_2$ and MoSe$_2$ and $\theta$ the twist angle between the two layers.
\\
Excitons confined in the moiré lattice obey the following Hamiltonian $H$
\begin{equation}
H(x,p_x)=\frac{p_x^2}{2m}+V_0 \sin^2{(Qx)}
\end{equation}

The Schrödinger equation is then
\begin{equation}
H(x,p_x)\phi^{n}_q(x)=E^{n}_q\phi^{n}_q(x)
\end{equation}
The eigenvectors $\phi^{n}_q(x)$ are Bloch wave functions, product of a plane wave $e^{iqx}$ with wavevector $q$ and a function $u^{n}_q(x)$, with the same periodicity as $V(x)$. $E^{n}_q$ are eigenenergies, i.e. the energies of the bands with index $n$ associated to the potential $V(x)$.

\subsection{Wannier functions}
Excitons localized in moiré lattice sites are best described by Wannier wavefunctions, which are exponentially localized around the sites and defined by
\begin{equation}
w^{n}_j(x)=\sqrt{\frac{a_m}{2\pi}}\int_{-\pi/a_m}^{\pi/a_m}\phi^{n}_q(x)\exp{(-ix_jq)}dq
\end{equation}

\subsection{Non interacting Hamiltonian}

In second quantization, the non interacting Hamiltonian reads
\begin{align}
h&=\sum_n \sum_{j,l} J_n(j-l) \hat{a}^{n\dagger}_l\hat{a}^{n}_j \\
J_n(j-l)&=\frac{a}{2\pi}\int_{-\pi/a_m}^{\pi/a_m}E^{n}_q  \exp{[i(j-l)aq]} dq
\label{tunnel}
\end{align}
where $\hat{a}^{n}_j$ annihilates an exciton with the Wannier wavefunction $w^{n}_j(x)$ on site $j$ with band index $n$.

The Hamiltonian $h$ describes the hopping from a site $j$ to a site $l$, with an amplitude $J_n(j-l)$ that depends on the band index $n$ and on the distance $|j-l|a$ between the two sites. $J_n(j-l)$ is in fact equal to the matrix element of the Hamiltonian between two Wannier functions
\begin{equation}
 J_n(j-l)=\int w^{n*}_j(x)\left[-\frac{\hbar^2}{2m_x}\frac{d^2}{dx^2}+V(x)\right]w^{n}_l(x)dx
\end{equation}

As Wannier functions decay exponentially when $|x-x_j|$ is large, their overlap drops rapidly when $|j-l|$ increases, so that we only take into account nearest neighbor hopping $J_n(1)=-t_n$, as represented by the sum $<j,l>$ in the final expression of the Hamiltonian
\begin{equation}
H=\sum_n \sum_{<j,l>} -t_n \hat{a}^{n\dagger}_l\hat{a}^{n}_j
\end{equation}

\subsection{Wannier energies}
$J_n(0)$ provides the energy associated with a given Wannier function localized on a site $j$, $w^{n}_j(x) \equiv  w^{n}_0(x)$.
For a given site $j$, there are $n$ Wannier states with energies $J_n(0)$ that correspond to Bloch bands energies averaged over the first Brillouin zone

\begin{equation}
J_n(0)=\frac{a}{2\pi}\int_{-\pi/a_m}^{\pi/a_m}E^{n}_q dq
\end{equation}

To draw Fig.3 in main text, we calculated the Wannier energy levels using the above relation with a fixed $V_0$= 30 meV and varying $a_m$ from 6 nm to 11 nm.
The energies $E^{n}_q $ of the Bloch functions are obtained by solving the 1D Schrödinger equation for a sinusoidal potential.
Fig.SI.6 shows, for example, the energies of confined states for a moiré period of 7 nm, 9 nm and 11 nm.  A second confined level appears around 11 nm period as also shown in Fig.3 of the main text.\vspace{0.3cm}

\centerline{\includegraphics[width=0.6\columnwidth]{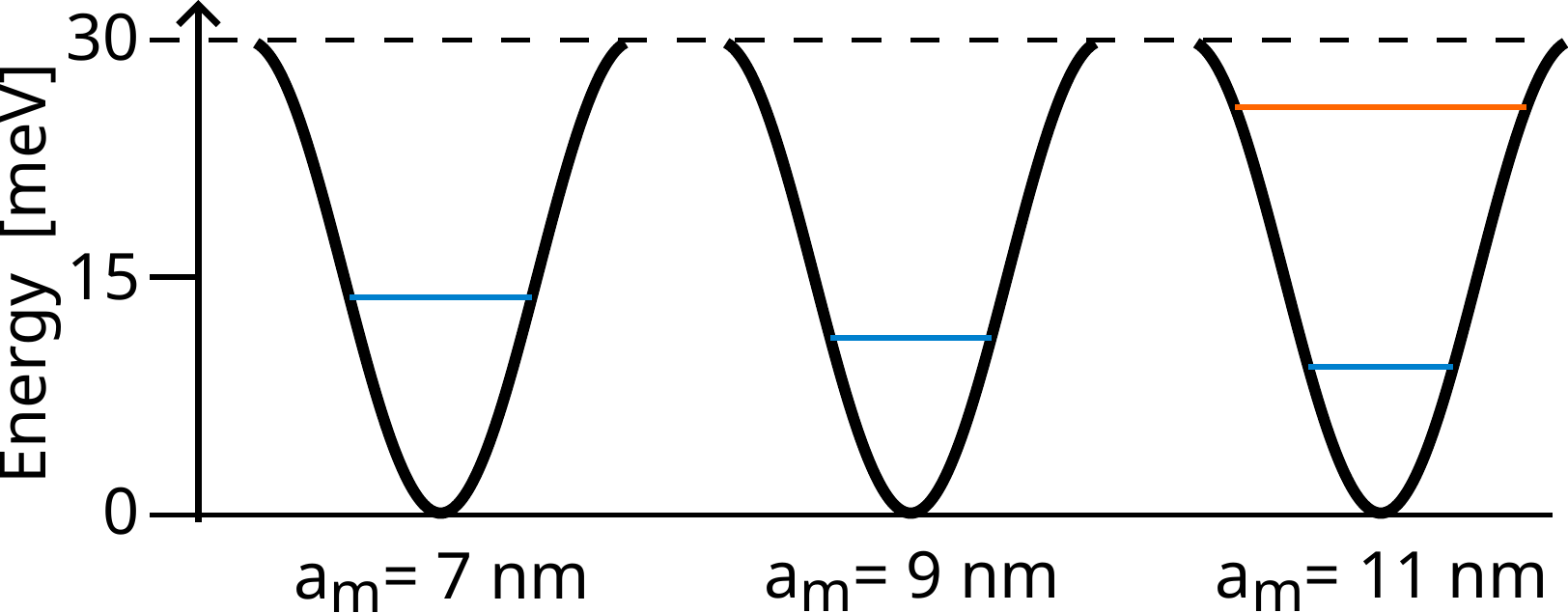}}
{\textbf{Fig. SI.6 Wannier energy levels.} Confined states energies for a moiré period equal to 7 nm, 9 nm and 11 nm, and a potential depth of 30 meV (from left to right).}

\end{document}